\documentclass[A4]{aa}
\usepackage{psfig} 
\usepackage{graphicx}
\def\M21{\hbox{Mrk\,421} }
\def\xmm{\hbox{$XMM-Newton$}}
\def\etal{et al. }

\def\ltsima{$\; \buildrel < \over \sim \;$}
\def\simlt{\lower.5ex\hbox{\ltsima}}            
\def\gtsima{$\; \buildrel > \over \sim \;$}
\def\simgt{\lower.5ex\hbox{\gtsima}}            

\newcommand{\eqref}[1]{(\ref{#1})}
\newcommand{\dsfrac}[2]{\displaystyle{\frac{#1}{#2}}}
\def\difd{{\rm d}}
\def\zb{z_{\rm b}}
\def\zf{z_{\rm f}}
\def\yb{y_{\rm b}}
\def\yf{y_{\rm f}}
\def\gb{\Gamma_{\rm b}}
\def\gf{\Gamma_{\rm f}}
\def\kf{K_{\rm f}}
\def\kb{K_{\rm b}}
\def\Tend{T_{\rm end}}
\def\Te{T_{\rm e}}
\def\Ti{T_{\rm i}}
\def\xb{x_{1{\rm b}}}
\def\xf{x_{1{\rm f}}}
\def\xbb{x_{2{\rm b}}}
\def\xff{x_{2{\rm f}}}
\def\MA04{MAMB04}
\newcommand{\SF}[2]{{\bf S}#1-{\bf F}#2}

\begin{document}

\title{Which physical parameters can be inferred \\ 
       from the emission variability of relativistic jets?}
\author{P.\,Mimica\inst{1}, M.A.\,Aloy\inst{1,2}, E.\,M\"uller\inst{1} 
        and W.\,Brinkmann\inst{3}}
\offprints{PM, e-mail: pere@mpa-garching.mpg.de}
\institute{Max--Planck--Institut f\"ur Astrophysik,
            Postfach 1312, D-85741 Garching, Germany
\and       Departamento de Astronom\'{\i}a y Astrof\'{\i}sica, 
           Universidad de Valencia, 46100 Burjassot, Spain
\and       Max-Planck-Institut f\"ur extraterrestrische Physik,
           Postfach 1603, D-85740 Garching, Germany}
\date{Received ?; accepted ?}
 \abstract{
   We present results of a detailed numerical study and theoretical
   analysis of the dynamics of internal shocks in relativistic jets
   and the non-thermal flares associated with these shocks. In our
   model internal shocks result from collisions of density
   inhomogeneities (shells) in relativistic jet flows. We find that
   the merged shell resulting from the inelastic collision of shells
   has a complicated internal structure due to the non-linear dynamics
   of the interaction.  Furthermore, the instantaneous efficiency for
   converting kinetic energy into thermal energy is found to be almost
   twice as high as theoretically expected during the period of
   significant emission.  The Lorentz factors of the internal shocks
   are correlated with the initial inertial masses of the shells. 
   
   Because of the complexity of the non-linear evolution the
     merged shell becomes very inhomogeneous and simple {\em one-zone}
     models are inadequate to extract physical parameters of the
     emitting region from the resulting light curves. In order to
     improve on these one-zone approximations, we propose a novel way
   of analyzing the space-time properties of the emission. Based on
   these properties we construct an analytic model of non-thermal
   flares which can be used to constrain some (unobservable) physical
   parameters of the internal shocks. These are the ratio of the
   Lorentz factors between the forward and the reverse shock (caused
   by the shell collision), and the shell crossing times of these
   shocks. The analytic model is validated by applying it to the
   synthetic light curves computed from our models. It can equally
   well be applied to observations.

  }
\titlerunning{Which physical parameters of relativistic jets can be inferred?}
\authorrunning{Mimica et. al}
\maketitle

\section{Introduction}
\label{intro}

BL Lac objects are a class of active galactic nuclei (AGN) which show
the most rapid variability of all AGNs. In the X-ray frequency range
they display flares whose duration in the observer frame is usually of
the order of one day (\cite{MA99} 1999, \cite{TA00} 2000, \cite{KA01}
2001), as further confirmed by the longterm observation of Mrk~501,
PKS\,2155-304, and Mrk\,421 (\cite{TA01} 2001, \cite{T01} 2001). With
the highly improved sensitivity of \xmm, variability on time scales
down to a few kiloseconds could be studied in Mrk\,421 (\cite{BR01}
2001, 2003) and only recently, from an \xmm observation of Mrk\,421 in
a very high state, the spectral evolution of the object down to time
scales of $\geq$ 100 sec could be followed (\cite{BR05} 2005).

The internal shock scenario (\cite{RM94} 1994) is usually invoked in
order to explain the variability of blazars (\cite{SP01} 2001,
\cite{BW02} 2002). These shocks are typically assumed to arise due to
the collisions of density inhomogeneities within the relativistic
blazar jets. Recently, one--dimensional (\cite{KI04} 2004) and
two--dimensional (\cite{MA04} 2004; thereafter \MA04) simulations of
such internal shocks have been performed both showing that the
evolution of internal shocks is considerably more complicated than
what has been concluded from approximate analytic considerations.  As
a result of the complexity of the non-linear evolution arising after a
two-shell interaction, the merged structure becomes very inhomogeneous
and simple {\em one-zone} models are inadequate to extract physical
parameters of the emitting region from the resulting light curves. In
particular, we will show that the rest-mass density of the emitting
shell can hardly be inferred from fits to the synthetic light curves.
Assuming an approximately uniform proportionality between the
rest-mass energy and the magnetic field energy the magnetic field
strength might hardly be inferred from a flare light curve, too.
Should the evolution of real shell interactions be as complex, the
parameters obtained from {\em one-zone} model fits to the spectra will
be equally inaccurate as those obtained from {\em one-zone} model fits
of the light curves.

In this paper we both extend our simulations of internal shocks and
perform a systematic study of the influence of the fluid properties on
the formation of flares, on their duration and temporal profiles, and
develop an analytic model linking the fluid properties to the observed
flare properties. One of the advantages of our analytic model is that
it can be validated by comparing its predictions to the actual
physical conditions obtained from our relativistic hydrodynamic
simulations. This validation will probe to be very important
considering the disparity of physical values predicted by other
analytic models for single flares in certain blazars (see
\S\,\ref{sec:discussion}).

In section \S\,\ref{sec:internal_shocks} we discuss the internal shock
scenario as a model which explains the flare emission and we
  summarize our previous findings (\MA04) regarding the hydrodynamic
  evolution of internal shocks. The numerical simulations are
described in Section \S\,\ref{sec:numsim}. Finally, in
\S\,\ref{sec:analytic} we present an analytic model which can be used
to infer physical parameters of the fluid flow from either synthetic
or observed flare light curves (\S\,\ref{sec:fits} and
\ref{sec:spacetime}).

\section{Internal shock scenario}
\label{sec:internal_shocks}

The internal shock scenario assumes the existence of blobs of matter
moving with different velocities along the jet, presumably emitted by
the intermittently working central engine. Two blobs will interact
after a certain time depending on their initial relative
velocity. Once the interaction of the shells starts two internal
shocks form, one propagating into the slower shell (forward shock) and
another one into the faster shell (reverse shock). The interaction of
the blobs is modeled as a collision of two homogeneous shells
(\cite{SP01} 2001, \cite{TA03} 2003), and also directly simulated
(\MA04). These models shells are assumed to have sharp edges (density
discontinuities) separating them from the background fluid.

The present work relies on the physical model developed by \MA04. In
that model the comoving magnetic field in the region behind the shocks
is assumed to be randomly oriented in space. Its energy density is
proportional to the internal energy density of the shocked fluid. The
non-thermal particles simulated within the model are accelerated at
the shock front to very high Lorentz factors whereby they radiate
synchrotron radiation.

The observed light curve from a two-shell collision depends on the
position along the jet at which the interaction occurs as well as on
the details of the evolution. As we discussed in \MA04, the
hydrodynamic evolution of the shells begins already before the
collision time predicted by the analytic models (e.g. \cite{SP01}
2001) and its most important features are (\MA04):

\begin{itemize}
\item The front (with respect to the direction of motion)
discontinuity of each shell decays into a bow shock, a contact
discontinuity and a reverse shock.
  
\item The back discontinuity of each shell develops into two
rarefactions. The first one connects the still unperturbed material
inside the shell with a contact discontinuity separating shell matter
and external medium. The second rarefaction connects the contact
discontinuity with the external medium.
\end{itemize}

The pre--collision evolution is quantitatively similar independent of
whether the edges of the colliding shells are modeled as sharp
discontinuities or by a smooth transition layer.  The pre-collision
evolution of the shells is important as the internal energy of the
front part of the shells rises due to the supersonic propagation of
the shells. Hence, when they actually start colliding the conditions
inside the shells are different from those at ejection from the
central engine, i.e., the conditions at the contact surfaces of the
two shells are different from those of the initial shells.  Once the
shells start to collide the evolution consists of three basic stages:

\begin{itemize}
\item The pressure in the region between the two shells rises rapidly
(compared to the light crossing time of the shell). The front and the
reverse shock propagate into the slower and faster shell,
respectively.

\item The shocks heat up the cold shell material. While the heated
region expands the pressure decreases slightly.

\item The front shock breaks out of the slower shell, enters the less
dense external medium, and thus accelerates. The pressure decreases
more rapidly during this stage because of the faster expansion.
\end{itemize}

\section{Numerical simulations}
\label{sec:numsim}
  
In our previous paper we simulated shell collisions in two spatial
dimensions. The results showed that the lateral expansion is
negligible during the collision of aligned shells (\MA04, Fig.\,4).
Thus, all essential features of aligned shell interactions can be
captured using 1D simulations. This allowed us to compute seven
high-resolution 1D models all having the same initial shell velocities
and the same shell geometries. The models include one where the rest
mass densities of both shells are identical (\SF{10}{10}), two where
both shells have approximately the same conserved mass (\SF{10}{07},
\SF{14}{10}), and one where both shells possess the same kinetic
energy (\SF{10}{05}). The latter three models were also simulated by
exchanging the initial rest mass density of the colliding shells
(Tab.\,\ref{tab:1}).

  \begin{table*}
      \begin{center}
        \begin{tabular}{|c||c|c|c||c|c|c||c|c|}
          \hline
          sim. & $\dsfrac{\rho_1}{10^3\rho_{\rm ext}}$ & $\Gamma_1$ &
          $\dsfrac{{\cal M}_1}{10^3\rho_{\rm ext}V}$ &
          $\dsfrac{\rho_2}{10^3\rho_{\rm ext}}$ & $\Gamma_2$ & $\dsfrac{{\cal
          M}_2}{10^3\rho_{\rm ext}V}$ & $\gb^{\rm sim}$ & $\gf^{\rm sim}$ \\\hline\hline
          \SF{10}{10} & $1$ &$5$ & $5$ & $1$ &$7$ & $7$ & $5.70$ & $6.04$
          \\\hline
          \SF{10}{07} & $1$ &$5$ & $5$ & $0.7$ &$7$ & $4.9$ & $5.61$ & $5.97$
          \\\hline
          \SF{10}{05} & $1$ &$5$ & $5$ & $0.5$ &$7$ & $3.5$ & $5.52$ & $5.89$
          \\\hline
          \SF{10}{14} & $1$ &$5$ & $5$ &  $1.4$ & $7$ & $9.8$ & $5.78$ & $6.12$ \\\hline
          \SF{07}{10} & $0.7$ &$5$ & $3.5$ & $1$ &$7$ & $7$  & $5.79$ & $6.12$\\\hline
          \SF{05}{10} & $0.5$ &$5$ & $2.5$ & $1$ &$7$ & $7$  & $5.88$ & $6.20$\\\hline
          \SF{14}{10} & $1.4$ &$5$ & $7$ & $1$ &$7$ &$7$ & $5.61$ & $5.96$ \\\hline
        \end{tabular}
      \end{center}
      \caption{Overview of the shell properties (densities $\rho_1$
        and $\rho_2$, Lorentz factors $\Gamma_1$ and $\Gamma_2$, and
        inertial masses ${\cal{M}}_1$ and ${\cal M}_2$) of the seven
        1D models. The initial thickness of the shells is
        $10^{14}$\,cm, their volume is $V$, and their initial
        separation is $D_0 = 10^{14}$\,cm. The density of the external
        medium $\rho_{\rm ext} = 10^{-23}$\,g cm$^{-3}$, and its
        Lorentz factor $\Gamma_{\rm ext} = 2.9$. Initially $p/\rho c^2
        = 10^{-4}$ everywhere. The parameters of the type--E shock
        acceleration model (see \MA04 for details) are: $\alpha_{\rm
        e} = 10^{-2}$, $\gamma_{\rm min} = 30$, $\eta = 7\,10^3$. The
        last two columns give the Lorentz factor of the reverse
        ($\gb^{\rm sim}$) and forward ($\gf^{\rm sim}$) shocks,
        computed from the hydrodynamic simulations.
      }\label{tab:1}
  \end{table*}
      
The numerical grid consists of $10^4$ zones covering a physical domain
with a length of $5\,10^{15}$\,cm.  A re-mapping technique (see \MA04
for details) was applied in order to be able to follow the evolution
of the two shells initially separated by $10^{14}$\,cm up to distances
of $10^{17}$\,cm from the AGN engine. The re-mapping technique allowed
us to resolve spatial scales from $5\,10^{11}$\,cm to $10^{17}$\,cm.
      
The energy distribution of the non--thermal electrons is covered by 64
energy bins, and the number of frequencies at which we compute the
synchrotron radiation is $25$, logarithmically spanning the frequency
range from $10^{16}$ to $10^{19}$\,Hz. We use the type-E shock
acceleration model of \MA04.
      
The integration of the conservation laws of relativistic hydrodynamics
and the computation of the synchrotron radiation are performed using
\emph{RGENESIS}, which is an extension of the code \emph{GENESIS}
(\cite{AL99} 1999). The handling of the non--thermal particles within
\emph{RGENESIS} is described in detail in \MA04.

\subsection{Hydrodynamic evolution}
\label{sec:hydro}
        
\begin{figure*}
      \centering
      \includegraphics[scale=0.3]{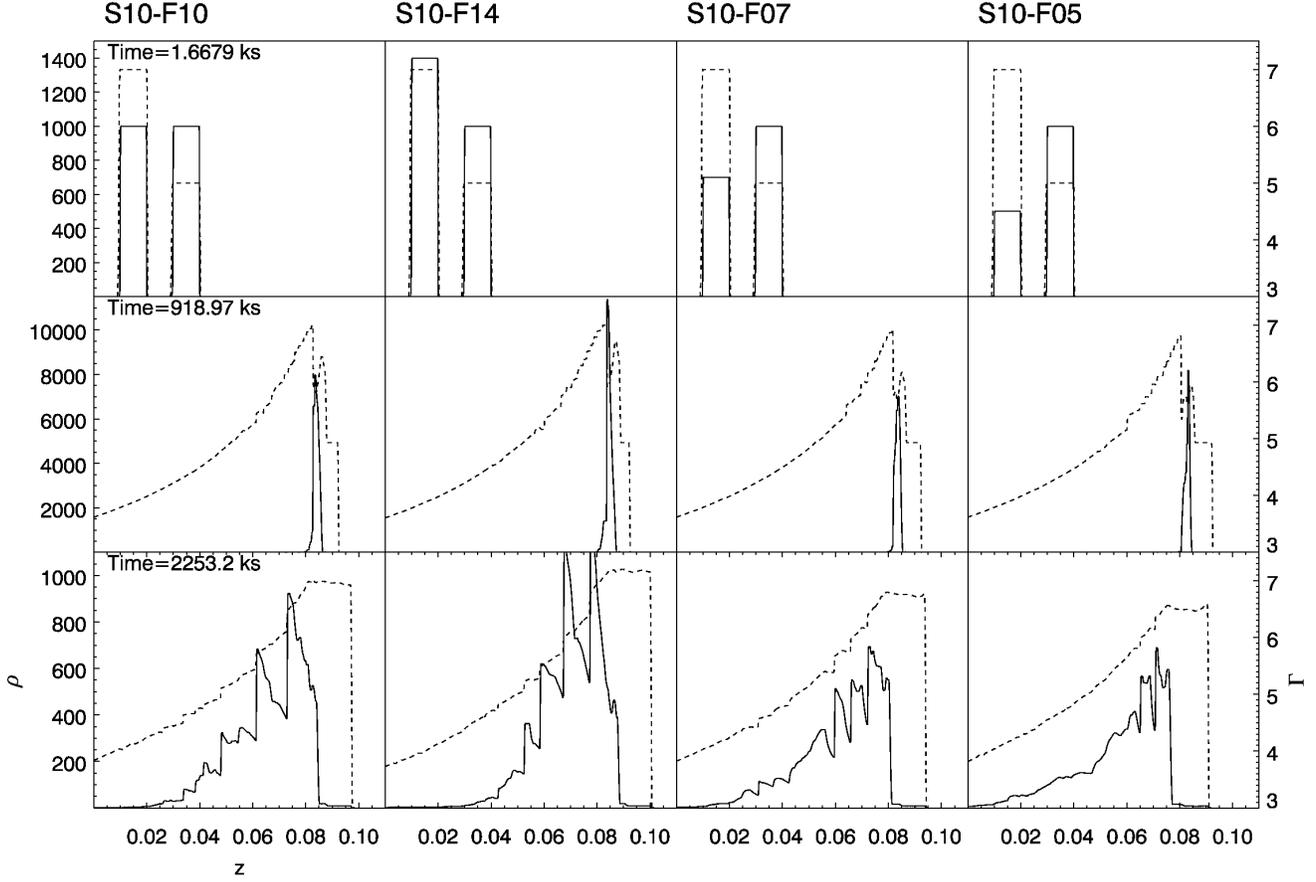}
      \caption{ Hydrodynamic evolution of the shell collision for
        models \SF{10}{10}, \SF{10}{14}, \SF{10}{07}, and \SF{10}{05}.
        The solid lines give the rest mass density in units of
        $\rho_{\rm ext}$. The dashed lines show the Lorentz factor of
        the fluid which is moving to the right. The origin of the
        z-axis corresponds to a distance of 0\,cm (top row),
        $2.65\,10^{16}$\,cm (middle row), and $6.61\,10^{16}$\,cm
        (bottom row), respectively. Times given in the leftmost panels
        of each row are measured in the laboratory frame.}
      \label{fig:1} 
    \end{figure*}

\begin{figure*}
      \centering
      \includegraphics[scale=0.3]{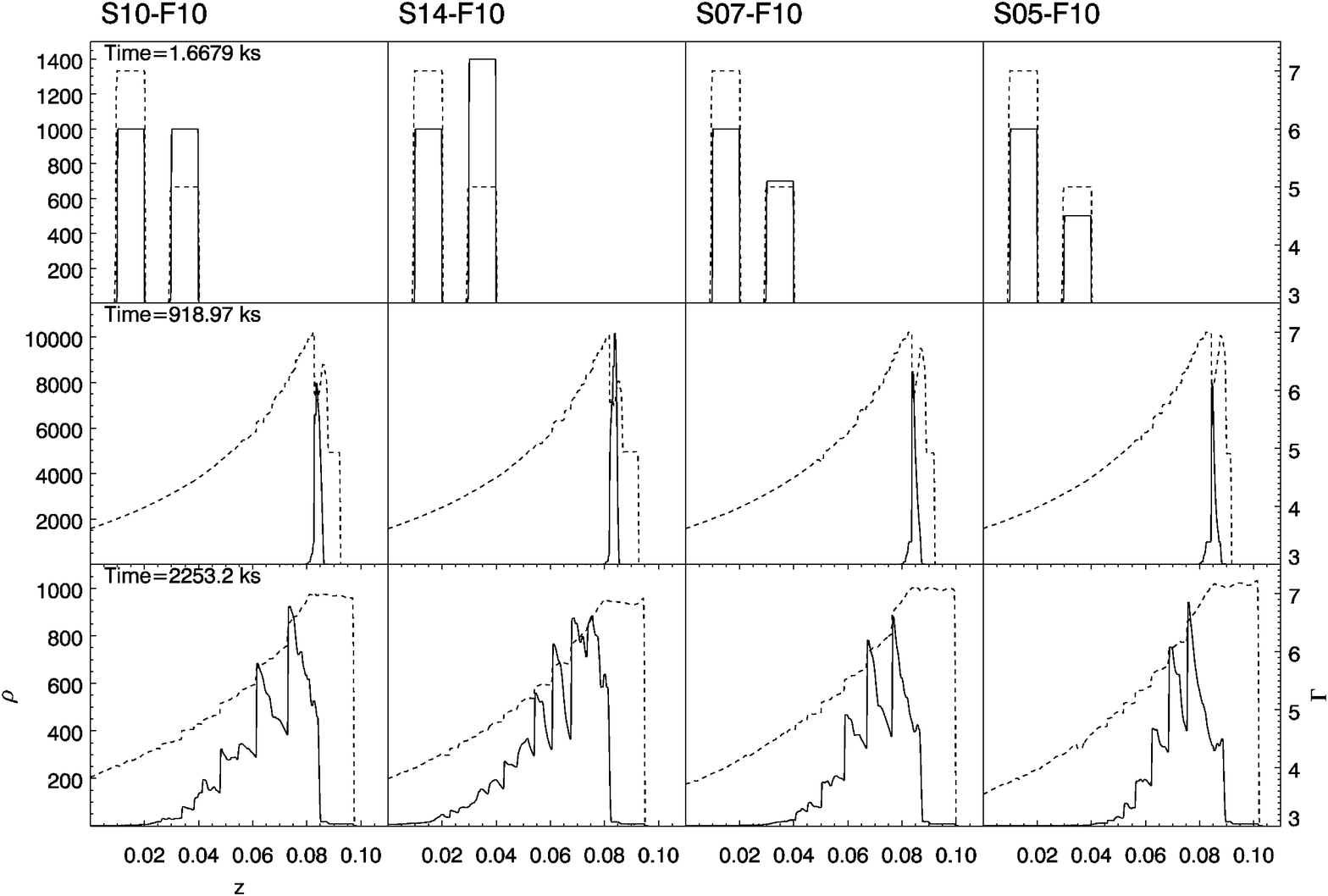}
      \caption{ Same as Fig.\,\ref{fig:1}, but for models \SF{14}{10},
        \SF{07}{10} and \SF{05}{10}. Model \SF{10}{10} is shown once
        again here to allow for an easier comparison. } 
      \label{fig:2}
    \end{figure*}

The main properties of the hydrodynamic evolution of our models are
summarized in Figs.\,\ref{fig:1} and \ref{fig:2} showing the initial
models (top row), the formation of the internal shocks (middle row),
and the final state (bottom row).

The evolution depends on the density of the shells in the following
way: the higher the density of the faster shell, the higher is the
resulting density peak (middle rows of Figs.\,\ref{fig:1} and
\ref{fig:2}), and the larger are the density variations within the
merged shell (bottom rows in Figs.\,\ref{fig:1} and \ref{fig:2}).  The
shock Lorentz factors, as computed from the simulations ($\gb^{\rm
sim}$, $\gf^{\rm sim}$), increase with the rest mass density of the
faster shell.
    
Kino et al.\,(2004) obtain a multi--peaked final density distribution,
which they attribute to the rarefaction waves forming when the
internal shocks cross the edges of the shells. In our models the
rarefaction waves cause the Lorentz factor to decrease towards the
rear part of the merged shell (bottom rows in Figs.\,\ref{fig:1} and
\ref{fig:2}).  However, the direct comparison with \cite{KI04}\,(2004)
is difficult, as their results are computed in the comoving frame of
the contact discontinuity between the two shells, whereas our results
are shown in the laboratory frame attached to the central engine.
Despite of this difficulty, there is a robust result common to every
simulation modeling two-shell interactions, namely the formation of a
very inhomogeneous merged structure as a result of the non-linear
evolution of the system. This heterogeneity flaws simple {\em
  one-zone} models assuming homogeneous conditions in the emitting
regions.

\subsection{Energy conversion efficiency}
\label{sec:convef}
      
Assuming an inelastic collision of two shells, the following
conversion efficiency $\epsilon$ of bulk kinetic energy into internal
energy can be derived from the conservation of energy and momentum
(\cite{KPS97}~1997; \cite{DM98}~1998):

      \begin{equation}\label{eq:epsilon}
        \epsilon=1-\dsfrac{({\cal M}_1+{\cal
            M}_2)\Gamma_m}{{\cal M}_1\Gamma_1+{\cal M}_2\Gamma_2}\, ,
      \end{equation}
%
where ${\cal M}_i=\rho_i \Gamma_i V$ is the total inertial mass of the
shell $i$ ($i=1, 2$), and where the Lorentz factor of the merged shell
is given by
%
      \begin{equation}\label{eq:Gamma_m}
        \Gamma_m=\sqrt{\Gamma_1\Gamma_2\dsfrac{{\cal M}_1\Gamma_1+{\cal
            M}_2\Gamma_2}{{\cal M}_1\Gamma_2+{\cal M}_2\Gamma_1}}\, .
      \end{equation}
        
The (time independent) efficiency given in Eq.\,\eqref{eq:epsilon}
corresponds to a situation where two incompressible shells collide
inelastically and instantaneously.  However, as we are modeling the
shells as compressible fluids, we have to consider the temporal
evolution of the energy conversion efficiency, and hence use the
following generalized definition
%
      \begin{equation}\label{eq:epsilont}
        \epsilon(T)\equiv1-\dsfrac{\int \Gamma^2(T, z)\rho(T, z)\difd
          z}{\int \Gamma^2(0, z)\rho(0, z)\difd z}\, ,
      \end{equation}
%
where $\rho(T, z)$ and $\Gamma(T, z)$ are the rest mass density and
the Lorentz factor of the fluid at a time $T$ at position $z$,
respectively. We point out that the instantaneous values of
$\epsilon(T)$ do not necessarily coincide with those given by
Eq.\,\eqref{eq:epsilon}, because during the hydrodynamic evolution the
fluid suffers multiple compressions and expansions changing the ratio
of internal energy and kinetic energy (which is assumed to be constant
in the simple model).
        
      \begin{figure}
        \centering
        \includegraphics[scale=0.31]{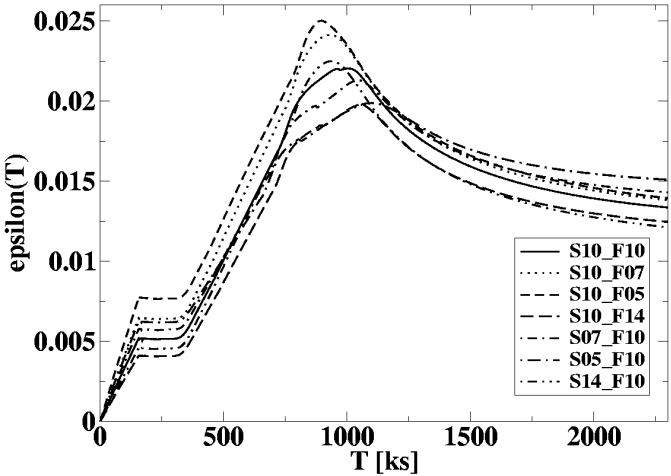}
        \caption{ Temporal evolution of the conversion efficiency
          defined by Eq.\,\eqref{eq:epsilont} for all models. $T$ is
          the time measured in the source frame.  }
        \label{fig:3}
      \end{figure}        
        
Fig.\,\ref{fig:3} shows the temporal evolution of conversion
efficiency $\epsilon(T)$ for all of our simulated models.  Initially,
the efficiency rises until about $150$\,ks, and then remains nearly
constant until $350$\,ks. The analysis of the hydrodynamic evolution
shows that this is due to the fact that the faster shell is heated due
to its interaction with the medium ahead of it. However, at $150$\,ks
the shell encounters a rarefaction created by the slower shell
reducing the amount of heating. At $330$\,ks the internal shocks form
and the efficiency starts rising again until it reaches a maximum at
about $1000$\,ks.  The subsequent decrease of the efficiency is caused
by the expansion of the fluid as the shocks break out of and
rarefactions form within the merged shell. The efficiency is generally
the higher the smaller is the rest-mass density of the faster shell
(Tab.\,\ref{tab:2}).  We note that the final values approach the
efficiency predicted by the conservation laws
(Eq.\,\eqref{eq:epsilon}).

      \begin{table}
          \begin{center}
            \begin{tabular}{|c||c|c|c|}
              \hline
              sim. & $\epsilon_{\rm max}$ & $\epsilon_{\rm fin}$ &
              $\epsilon_{\rm pred}$\\\hline\hline
              \SF{10}{10} & $0.0220$ & $0.0133$ & $0.0136$
              \\\hline
              \SF{10}{07} & $0.0241$ & $0.0137$ & $0.0140$
              \\\hline
              \SF{10}{05} & $0.0250$ & $0.0138$ & $0.0135$
              \\\hline
              \SF{10}{14} & $0.0198$ & $0.0123$ & $0.0125$  \\\hline
              \SF{07}{10} & $0.0210$ & $0.0142$ & $0.0125$  \\\hline
              \SF{05}{10} & $0.0200$ & $0.0149$ & $0.0109$  \\\hline
              \SF{14}{10} & $0.0225$ & $0.0121$ & $0.0140$  \\\hline
            \end{tabular}
          \end{center}
          \caption{ Maximum conversion efficiency $\epsilon_{\rm
            max}$, final conversion efficiency $\epsilon_{\rm fin}$,
            and theoretically predicted conversion efficiency
            $\epsilon_{\rm pred}$ (see Eq.\,\eqref{eq:epsilon}) for
            the seven simulated models. }
        \label{tab:2}  
      \end{table}
        
The time evolution of the conversion efficiency in our models is
qualitatively similar to that of \cite{KI04}(2004). However, due to
the fact that these authors computed the evolution in the rest frame
of the (relativistically moving) contact discontinuity, a quantitative
comparison with our results is presently impossible.

\subsection{Synthetic light curves}
\label{sec:flares}
      
      \begin{figure}
        \centering
        \includegraphics[scale=0.32]{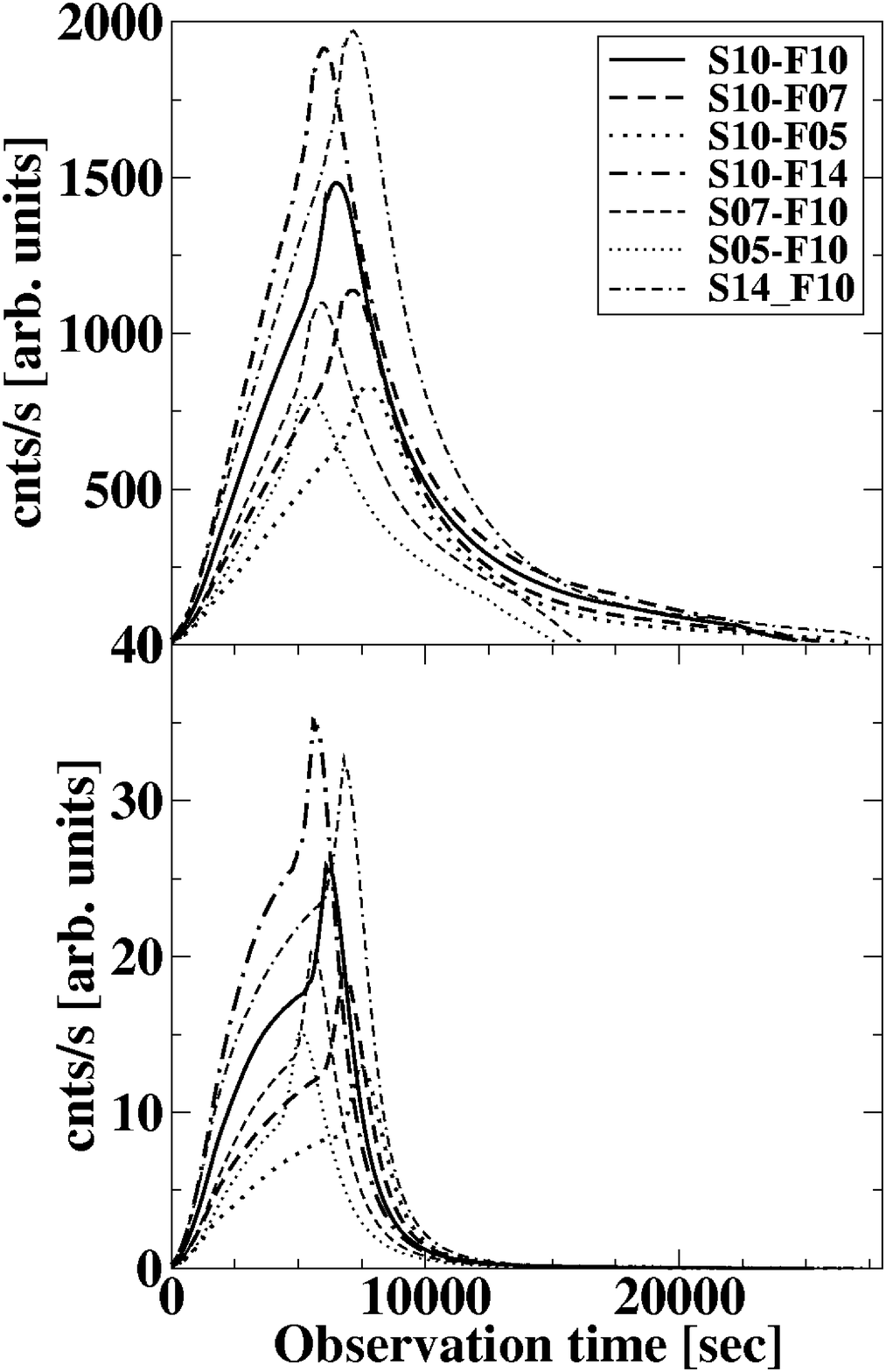}
        
        \caption{ Soft (upper panel; observed photon energies between
          $0.1$ and $1$\,keV) and hard (lower panel; $2$-$10$ keV)
          light curves for all of our seven models. A correlation
          between the total rest-mass of the shells and the peak
          photon counts is seen.  }
        \label{fig:4} 
     \end{figure}
      
The peak intensity of the light curves in both the soft
(Fig.\,\ref{fig:4}; upper panel; $0.1-1$\,keV) and the hard energy
band (Fig.\,\ref{fig:4}; lower panel; $2-10$\,keV) is correlated to
the total rest-mass of the shells (see Tab.\,\ref{tab:1}). This
correlation is slightly better in the wider hard energy band. We
further find that the peak of the flare occurs in the hard energy band
several hundreds of seconds before the peak in the soft band.
        
The shape of the flares is qualitatively similar for all models
indicating that the variation of the density ratio between shells
mostly tends to change the normalization of the flares.  Furthermore,
the large variations of the density (Figs.\,\ref{fig:1} and
\ref{fig:2} bottom rows) and of the specific internal energy are not
correspondingly do not change the shape of the light curves which are
rather smooth.  This implies that from the shape of the flare alone,
it is very hard to infer any information about the rest--mass density
or the specific internal energy of the emitting regions. This
restriction should be taken into account by any analytic model aiming
to infer physical parameters from light curves.
    
\subsection{Sensitivity to the initial conditions}

The light curves generated from our models are rather insensitive to
both the choice of density and pressure of the external medium
provided they are sufficiently small to be {\em inertially
  negligible}. If the density of the underlying medium is comparable
to that of the shells the evolution will be quantitatively different
(but qualitatively, the same Riemann structure will develop from every
edge of every shell).  We have limited the Lorentz factor of the
external medium to be smaller than that of the slowest shell. Its
value changes only the depth of the rarefaction trailing both shells
(the smaller the velocity of the jet, the smaller the pressure and the
density reached behind every shell).  However, this will not change
the light curves of our models as from rarefactions there is no
emission (because only shocked regions emit in our model).  The
complementary situation, i.e., one or both shells being slower than
the jet, we have not considered. This situation will correspond to a
couple of shells being slower than the underlying, light jet.  In such
a case, the Riemann structure will be qualitatively different.  The
rear edges of the shells will be heated because of the development of
shocks while the pull of the external medium will create rarefactions
from the forward edges of every shell. Nevertheless, this interesting
situation falls beyond the scope of the current work and will be
considered elsewhere. We point out that it is reasonable to expect a
low--density external medium, since the shells will clean up the
medium through which they propagate (note that the state left behind
every shell is less dense than that set up initially). This
rarefication process of the external medium is a robust feature
associated with the shell motion (see MAMB04).

\section{Analytic modeling of flares}
\label{sec:analytic}
      
In this section we discuss some analytic approaches to the modeling of
flares resulting from internal shocks, and most importantly which
physical parameters can be obtained by analyzing the shape and
duration of flares. We begin with general considerations of the
observations of distant unresolved sources, and then provide an
analytic model particularly suited to study the flares resulting from
collisions of shells in a relativistic jet.

\subsection{Computation of the radiation observed by a distant observer}
\label{sec:obs}
        
As the simplest case, we assume that we have a source of radiation
moving along the line--of--sight (LOS) towards a distant observer. We
introduce two inertial frames: the \emph{source frame} and the
\emph{observer frame}. The source frame is attached to the central
energy source. In this frame the physics of the source is best
described because time and space are independent coordinates. In the
source frame, we denote the position along the LOS by $z$, and the
time by $T$. The observer frame is located at a distance $z = z_{\rm
  obs}$ along the LOS. Except for a cosmological redshift factor, this
frame is equivalent to one attached to the Earth.  In this frame we
have chosen two different coordinate systems, the $(z_{\rm obs},
t_{\rm obs})$--system and the $(x,y)$--system. The second coordinate
system is best suited for describing the structure of the observations
in space time because the coordinates associated with this frame mix
the spatial ($z$) and temporal ($T$) coordinates of the source frame
in the following way:
%
        \begin{equation}\label{eq:x_coord}
          x=\dsfrac{cT-z}{\sqrt{2}}\, ,
        \end{equation}
%
        \begin{equation}\label{eq:y_coord}
          y=\dsfrac{cT+z}{\sqrt{2}}\, ,
        \end{equation}
%
where $c$ is the speed of light.
        
A process which takes place at time $T$ and position $z$ in the source
frame is observed in the observer frame at a time
%
        \begin{equation}\label{eq:tobs}
          t_{\rm obs} = T+\dsfrac{z_{\rm obs}-z}{c}
                      = \dsfrac{\sqrt{2}x}{c} + \dsfrac{z_{\rm obs}}{c}\, ,
        \end{equation}  
%
i.e., all points with the same $x$ coordinate will be observed
simultaneously. Hence, the \emph{total} radiation observed by a
distant observer at time $t$ is equal to the integral over $y$ of the
emitted radiation of all points in the source frame which have the
same $x=(ct-z_{\rm obs})/\sqrt{2}$ coordinate in the observer frame
($xy$--coordinate system), i.e.,
%
        \begin{equation}\label{eq:lc}
          I(t)=\int\difd y\, j(x, y)\, ,
        \end{equation}
%
where $I(t)$ is the intensity of the observed radiation at time $t$,
and $j(x, y)$ is the emissivity in the observer frame. This emissivity
is related to the emissivity at the corresponding time and place in
the source frame, i.e., $j(x, y)=j(z(x, y), T(x, y))$.
        
If a point source is moving with a velocity $\beta c$ then $z(T)=
z_0+\beta cT$, and its trajectory in $xy$-coordinates will be
%
        \begin{equation}\label{eq:trajectory}
          y(x)=\dsfrac{1+\beta}{1-\beta}x+\dsfrac{\sqrt{2}}{1-\beta}z_0\, .
        \end{equation}
Hence, an object moving with the speed of light is observed to
propagate parallel to the $y$ axis, while an object at rest in the
source frame is observed to move parallel to the line $y=x$.

Now we assume that an observer is located at $z_{\rm obs}=0$, and that
the emissivity of a point source is a function of the time $T$,
%
        \begin{equation}\label{eq:simple}
          j(z, T)=\left\{ \begin{array}{ll}
            I_0\delta(z-z_0-\beta cT) & {\rm if\,} 0\leq T\leq \Tend\\[4mm]
            0 & {\rm otherwise}
          \end{array}
          \right.
        \end{equation}
%
where $I_0$, $z_0$, $\beta$ and $\Tend$ are the normalization of the
emissivity, the initial source position, its velocity in units of the
speed of light, and the duration of its emission, respectively. In the
observer frame ($xy$-coordinates) the emissivity is then, using
\eqref{eq:x_coord}, \eqref{eq:y_coord}, and the properties of the
$\delta$-function,
%
        \begin{equation}\label{eq:simplexy}
          \begin{array}{cc}
            j(x, y)=&
                    \dsfrac{I_0\sqrt{2}}{1-\beta} \delta
                    \left[ y-\dsfrac{x(1+\beta)+
                           z_0 \sqrt{2}}{1-\beta} 
                    \right]\times
            \\
            & S(y; -x, \sqrt{2} c \Tend - x)
          \end{array}
        \end{equation}
%
where 
%
        \begin{equation}\label{eq:S}
          S(x;a,b)=\left\{ \begin{array}{ll}
            1 & {\rm if\,} a\leq x\leq b\\[4mm]
            0 & {\rm otherwise}
          \end{array}\right. \, .
        \end{equation}
%
Using now equation \eqref{eq:lc} we can compute the light curve
produced by a single emitting point source observed by a distant
observer:
%
        \begin{equation}\label{eq:simplelc}
          I(t)=\dsfrac{I_0\sqrt{2}}{1-\beta}S\left(\dsfrac{ct}{\sqrt{2}};
          \dsfrac{-z_0}{\sqrt{2}},\dsfrac{(1-\beta)c\Tend-z_0}{\sqrt{2}}
          \right)\, .
        \end{equation}
%
It follows from Eq.\,\eqref{eq:simplelc} that for the observer the
duration of the emission is given by
%
        \begin{equation}\label{eq:duration}
          t_{\rm total}=(1-\beta)\Tend\, ,
        \end{equation}
%
or, in the ultra-relativistic limit by $t_{\rm total} \approx \Tend /
(2\Gamma^2)$, where $\Gamma = 1/\sqrt{1-\beta^2}$ is the Lorentz
factor of the emitting point source.

This shows that the total time (measured in the source frame) during
which the source emits and its Lorentz factor are degenerate for a
distant observer, i.e., it is not possible to determine \emph{both}
the Lorentz factor \emph{and} the source frame time of an emission
process from the duration of the observed emission alone. This is a
major obstacle for any model which attempts to reconstruct, from an
observed flare, the sizes and the velocities of colliding shells which
produce that flare. Nevertheless, in the following sections we develop
an analytic model which can recover some of the properties of the
colliding shells, taking into account this limitation.

\subsection{Geometry of the emitting region}
\label{sec:geometry_xy}
  
Motivated by the results of numerical simulations (\MA04) the emitting
flare can be represented as a horn--shaped region in the observer
frame when using the $xy$-coordinates (Fig.\,\ref{fig:5}).  The two
curves $\yb(x)$ and $\yf(x)$ delimiting the horn--shaped region
correspond to the front and back edges of the merged shell,
respectively. They are approximated by second order polynomials
passing through the origin (Fig.\,\ref{fig:5}).  This leaves two free
parameters for each curve, which are determined by requiring that the
velocities of the front and back edges be $\beta_{\rm fi}$ and
$\beta_{\rm bi}$ at $T=0$, and $\beta_{\rm fe}$ and $\beta_{\rm be}$
at $T=\Te$, respectively.  $\Te$ is the time (measured in the source
frame) at which the emissivity from the shell interaction drops below
the observable level. By computing the equations of motion (in the
observer frame) $\zf(T)$ and $\zb(T)$ corresponding to $\yf(x)$ and
$\yb(x)$, respectively, one obtains the following equations for the
front edge,
%
      \begin{eqnarray}
        \yf(x)=4\gf^2
        x\left[\kf\gf^2\dsfrac{x}{\sqrt{2}\Te}+1-\dsfrac{1}{4\gf^2}
         \right]\label{eq:yf}\,
        ,\\
        \zf(T)=cT\left\{1-\left[\gf^2\left(1+\sqrt{1+\kf\dsfrac{T}{\Te}}
               \right)\right]^{-1}\right\}\label{eq:zf}\, ,
      \end{eqnarray}
%
and the back edge
%
      \begin{eqnarray}
        \yb(x)=4\gb^2
        x\left[\kb\gb^2\dsfrac{x}{\sqrt{2}\Te}+1-\dsfrac{1}{4\gb^2}
         \right]\label{eq:yr}\,
        ,\\
        \zb(T)=cT\left\{1-\left[\gb^2\left(1+\sqrt{1+\kb\dsfrac{T}{\Te}}
               \right)\right]^{-1}\right\}\label{eq:zr}\, ,
      \end{eqnarray}
%
where $\gf$, $\gb$, $\kf$, and $\kb$ are defined as
%
      \begin{eqnarray}
        \gf\equiv\left[2 \left(1-\beta_{\rm fi}\right)\right]^{-1/2}\, ,
        \label{eq:gf}\\
        \gb\equiv\left[2 \left(1-\beta_{\rm bi}\right)\right]^{-1/2}\, ,
        \label{eq:gr}\\
        \kf\equiv\left(\dsfrac{1-\beta_{\rm fi}}{1-\beta_{\rm fe}}
                  \right)^2-1\, ,
        \label{eq:kf}\\
        \kb\equiv\left(\dsfrac{1-\beta_{\rm bi}}{1-\beta_{\rm be}}
                 \right)^2-1\, .
        \label{eq:kr}
      \end{eqnarray}
%
      We point out that \emph{only} in the ultra-relativistic case
      (Lorentz factors$\simgt 2$, $\,\beta$ close to $1$) $\gf$ and
      $\gb$ approach the Lorentz factors corresponding to the initial
      velocities of the front and back edges which in turn can be
      identified with the Lorentz factors of the forward and reverse
      shocks, respectively. However, as we are usually dealing with
      ultra-relativistic velocities in our model, there arises no
      significant error when substituting \eqref{eq:gf} and
      \eqref{eq:gr} by the corresponding Lorentz factors in equations
      (\ref{eq:yf})-(\ref{eq:zr}). In this case $\kf$ and $\kb$ are
      given by
%
      \begin{eqnarray}
        \kf=\left(\dsfrac{\Gamma_{\rm fe}}{\gf}\right)^4-1\, \label{eq:kfu}\\
        \kb=\left(\dsfrac{\Gamma_{\rm re}}{\gb}\right)^4-1\, \label{eq:kru}
      \end{eqnarray}
%
where $\Gamma_{\rm fe}$ and $\Gamma_{\rm be}$ are the Lorentz factors
of the front and back edge at $T=\Te$, respectively.

\subsection{Light curve of a horn--shaped segment}
\label{sec:horn_seg}
      
      \begin{figure}
        \centering
        \includegraphics[scale=0.42]{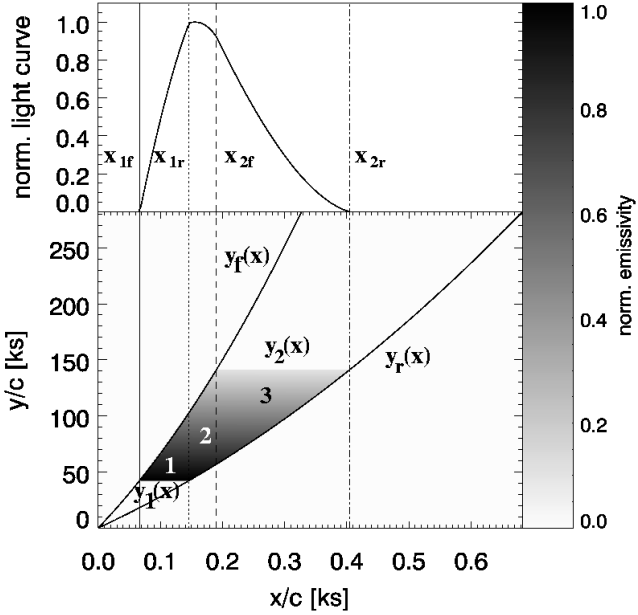}
        \caption{ Light curve (upper panel) and the observer
          emissivity distribution in $xy$--coordinates (lower panel)
          for a horn--shaped segment with parameters $\gb=8$, $\kb=4$,
          $\gf=12$, $\kf=3$, $T_1=30$\,ks, $T_2=100$\,ks,
          $\Te=200$\,ks, $\eta=0.1$, and $\theta=1$. The curves
          delimiting the horn--shaped region corresponding to the front
          and back edges of the merged shell are denoted by $\yf(x)$
          and $\yb(x)$, respectively. The edges of the segment
          corresponding to $T_1$ and $T_2$ are indicated by $y_1(x)$
          and $y_2(x)$, respectively.  The vertical lines correspond
          to the points $x_{\rm 1f}$ (solid), $x_{\rm 1b}$ (dotted),
          $x_{\rm 2f}$ (dashed), and $x_{\rm 2b}$ (dot-dashed),
          respectively. The numbers inside the horn--shaped area denote
          regions where the emissivity is integrated using different
          upper and lower limits.  }
        \label{fig:5} 
      \end{figure}
      
In order to compute the light curve emitted by the horn--shaped region
(Fig.\,\ref{fig:5}), we note that significant emission occurs only
after the actual interaction of the shells starts and that this
emission only lasts until the internal shocks reach the edge of the
merged shell which terminates the shell interaction. Hence, one needs
to calculate only the light curve produced by a segment of the
horn--shaped region defined by two times, $T_1$ (start of the
interaction) and $T_2$ (termination of the interaction). We assume
an emissivity distribution in the source frame of the form
%
      \begin{equation}\label{eq:jzT}
        \begin{array}{rl}
          j(z, T)=&
          j_0\left[1+\dsfrac{T^\theta-T_1^\theta}{T_2^\theta-T_1^\theta}
          (\eta-1)\right] \times \\[5mm]
          &S(T;T_1,T_2)S\left[z;\zb(T),\zf(T)\right]\, ,
        \end{array}
      \end{equation}
%
where $\theta$ is a power law index describing the temporal change of
the emissivity, and $\eta$ is the ratio of the final $j(T_2)$ and the
initial $j_0 \equiv j(T_1)$ emissivities.  The interval function
$S(T;T_1,T_2)$ (\eqref{eq:S} appearing in the above expression will be
dropped in the following discussion.

Substituting \eqref{eq:x_coord} and \eqref{eq:y_coord} into
\eqref{eq:jzT}, the emissivity in the observer frame
($xy$--coordinates) becomes
%
      \begin{equation}\label{eq:jxy}
        j(x, y)=\dsfrac{j_0}{T_2^\theta-T_1^\theta}\left[T_2^\theta-
          \eta T_1^\theta+\dsfrac{(x+y)^\theta}{(\sqrt{2}c)^\theta}
         (\eta-1)\right]\, .
      \end{equation}
%
The segment of the horn-shaped region is bound by the two parabolae
$\yf(x)$ and $\yb(x)$ in $x$--direction (see
\S\,\ref{sec:geometry_xy}), and by the following two straight lines in
$y$--direction (Fig.\,\ref{fig:5}),
%
      \begin{eqnarray}
        y_1(x)=\sqrt{2}cT_1-x\, ,\\
        y_2(x)=\sqrt{2}cT_2-x\, .
      \end{eqnarray}
%
The light curve $I(x)$ at a given $x$--coordinate is obtained by
integrating the emissivity distribution at that $x$--coordinate in
$y$--direction. According to our model the integration limits depend
on the value of the $x$--coordinate, and three regions can be
distinguished (Fig.\,\ref{fig:5}).  In region\,1 ($\xf \leq x \leq
\xb$) the integration extends from $y_1(x)$ to $\yf(x)$, in region\,2
($\xb < x \leq \xff$) from $\yb(x)$ to $\yf(x)$, and in region\,3
($\xff < x \leq \xbb$) from $\yb(x)$ to $y_2(x)$ (see
Fig.\,\ref{fig:5}).  The coordinates $\xb$, $\xf$, $\xbb$, and $\xff$
are given by
%
      \begin{eqnarray}
        \xb=\dsfrac{\sqrt{2}cT_1}{2\gb^2}\dsfrac{1}{1+\sqrt{1+\kb
            T_1/T_e}}\, ,\\
        \xf=\dsfrac{\sqrt{2}cT_1}{2\gf^2}\dsfrac{1}{1+\sqrt{1+\kf
            T_1/T_e}}\, ,\\
        \xbb=\dsfrac{\sqrt{2}cT_2}{2\gb^2}\dsfrac{1}{1+\sqrt{1+\kb
            T_2/T_e}}\, ,\\
        \xff=\dsfrac{\sqrt{2}cT_2}{2\gf^2}\dsfrac{1}{1+\sqrt{1+\kf
            T_2/T_e}}\, ,
      \end{eqnarray}
%
where we assume $\xb < \xff$ (see next section for discussion). Thus,
$\xf < \xb < \xff < \xbb$. Defining two auxiliary functions
%
      \begin{equation}\label{eq:g}
        g(\xi,K,\Gamma)=4\Gamma^2(\Gamma^2 K\xi^2+\xi)\, ,
      \end{equation}
%
and
%
      \begin{equation}\label{eq:H}
        \begin{array}{l}
          H(A, B, \tau_1, \tau_2, \theta, \eta)=
          \dsfrac{1}{\tau_2^\theta-\tau_1^\theta}\times\\[5mm]
          \, \, \, \, \, \, \, \, 
          \left[(\tau_2^\theta-\eta\tau_1^\theta)(B-A)+
            \dsfrac{(\eta-1)(B^{\theta+1}-A^{\theta+1})}{\theta+1}\right]
        \end{array}
      \end{equation}
%
allows us to write the light curve of a segment of the horn-shaped
region in a very compact way,
%
      \begin{equation}\label{eq:seglc}
        \begin{array}{l}
          I(x)=j_0\sqrt{2}c\Te\times \\[5mm]
          \, \, \, \, \, \, \, \, \left\{\begin{array}{ll}
          H\left[\dsfrac{T_1}{\Te},g_{\rm f}(x),
            \dsfrac{T_1}{\Te},\dsfrac{T_2}{\Te},\theta,\eta\right] & 
          {\rm if}\, \xf\leq x\leq \xb\\[4mm]
          H\left[g_{\rm b}(x),g_{\rm f}(x),
            \dsfrac{T_1}{\Te},\dsfrac{T_2}{\Te},\theta,\eta\right] & 
          {\rm if}\, \xb< x\leq \xff\\[4mm]
          H\left[g_{\rm b}(x),\dsfrac{T_2}{\Te},
            \dsfrac{T_1}{\Te},\dsfrac{T_2}{\Te},\theta,\eta\right] & 
          {\rm if}\, \xff< x\leq \xbb
          \end{array}
          \right.
        \end{array}
      \end{equation}
%
where $g_{\rm f}(x)$ and $g_{\rm b}(x)$ are given by
%
      \[
      g_{\rm f}(x)\equiv g\left(\dsfrac{x}{\sqrt{2}c\Te},\kf,\gf\right)\, ,
      \]
      \[
      g_{\rm b}(x)\equiv g\left(\dsfrac{x}{\sqrt{2}c\Te},\kb,\gb\right)\, .
      \]

The upper panel of Fig.\,\ref{fig:5} shows an example of a light curve
computed from Eq.\,\eqref{eq:seglc} for $\gb=8$, $\kb=4$, $\gf=12$,
$\kf=3$, $T_1=30$\,ks, $T_2=100$\,ks, $\Te=200$\,ks, $\eta=0.1$, and
$\theta=1$.

\subsection{Two-phase flare model}
\label{sec:twop}
      
As discussed in \S\,\ref{sec:internal_shocks} and by \MA04, the
interaction of two shells consists of three basic phases: a fast rise
of the pressure, a slow pressure decrease while the internal shocks
propagate through the shells, and finally the acceleration of the
front shock as it breaks out of the slower shell. In our analytic
model we simplify the situation by assuming only two phases, and hence
call it a \emph{two-phase flare model}. For the first phase ($0\leq
T\leq \Ti$) we assume that the initial and final values of the
emissivity are $j_0$ and $\eta_1 j_0$, respectively. In the second
phase ($\Ti \leq T\leq \Te$) the emissivity is assumed to abruptly
jump to a value $\eta_2 j_0$, and then to decrease to zero. The abrupt
jump is motivated by the fact that when the forward shock, created by
the interaction of the shells, is about to break out of the slower
shell, it encounters the reverse shock produced by the pre--collision
hydrodynamic evolution of its leading edge.  The interaction of these
two shocks causes a fast rise both of the pressure and the density,
i.e., $\eta_2 > \eta_1$.
      
      \begin{figure}
        \centering
        \includegraphics[scale=0.42]{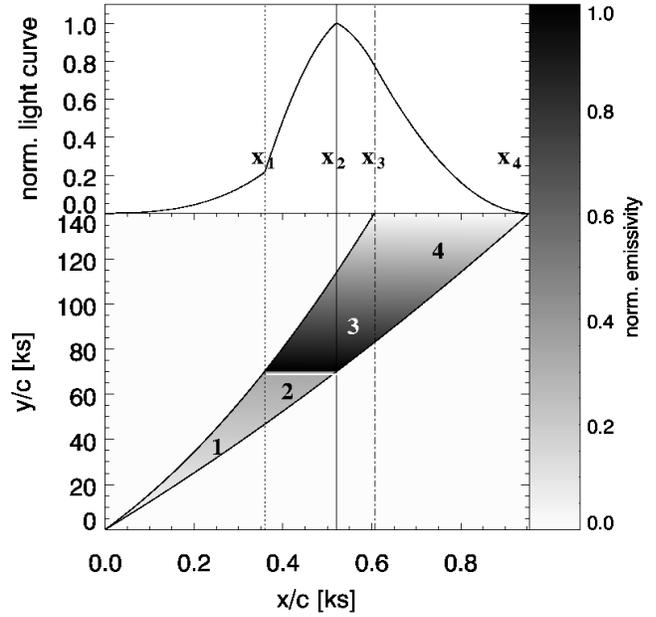}
        \caption{ Two--phase flare model with $\gb=5.5$, $\kb=1.1$,
          $\gf=6$, $\kf=4$, $\Ti=50$\,ks, $\Te=100$\,ks, $\theta_1=1$,
          $\theta_2=1$ and $\eta_1=100$, and $\eta_2=300$. The upper
          panel shows the normalized light curve, and the lower one
          the emissivity. The vertical lines denote the coordinates
          $x_1$ (solid), $x_2$ (dotted), $x_3$ (dashed), and $x_4$
          (dot-dashed line) which divide the emissivity distribution
          into four distinct regions labeled by numbers 1 to 4. The
          white thick horizontal line separates the two phases of the
          temporal evolution of the emissivity.  }
        \label{fig:6}
      \end{figure}
      
      \begin{figure}
        \centering
        \includegraphics[scale=0.42]{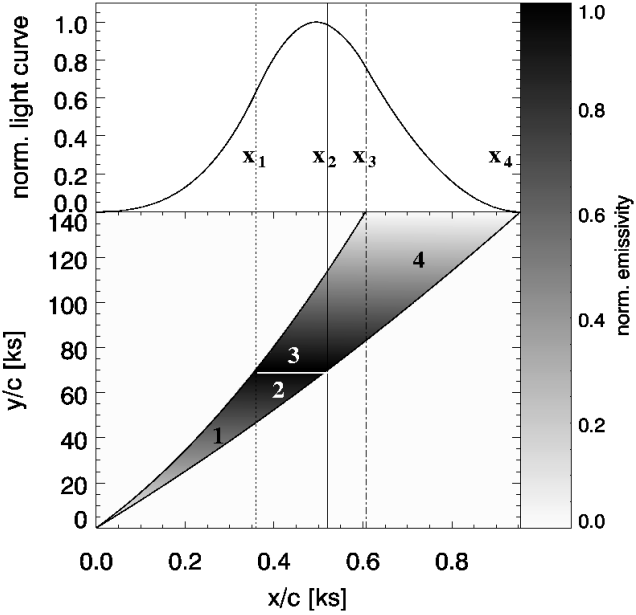}
        \caption{ Same as Fig.\,\ref{fig:6}, but for $\eta_2=101$. }
        \label{fig:7}
      \end{figure}
      
The light curve ${\cal I}(x)$ of the two--phase flare model is then
computed as follows: we assume that the emissivity $j(T)$ has the form
%
      \begin{equation}\label{eq:jT}
        j(T)=j_0\left\{
        \begin{array}{ll}
          1+\dsfrac{T^{\theta_1}}{\Ti^{\theta_1}}(\eta_1-1) & 
          {\rm if}\, 0\leq T\leq \Ti\\[5mm]
          \eta_2\left[1-\dsfrac{ T^{\theta_2}-\Ti^{\theta_2}}
                               {\Te^{\theta_2}-\Ti^{\theta_2}}
                \right] & {\rm if}\, 
          \Ti\leq T\leq \Te
        \end{array}
        \right.\, ,
      \end{equation}
%
and that the light curve in both phases (i.e., in both sub--segments
of the horn-shaped emission region; see (Fig.\,\ref{fig:6}) has the
form of the light curve $I(x)$ \eqref{eq:seglc} with common parameters
$\gb$, $\kb$, $\gf$, and $\kf$, but with the sub--segment specific
parameters according to the following substitutions:
      \begin{itemize}
      \item segment\, 1: \quad
        $j_0$, $T_1\rightarrow 0$, $T_2\rightarrow\Ti$,
        $\theta\rightarrow\theta_1$, $\eta\rightarrow\eta_1$
      \item segment\,2: \quad
        $j_0\rightarrow \eta_2 j_0$, $T_1\rightarrow
        \Ti$, $T_2\rightarrow\Te$, $\theta\rightarrow\theta_2$,
        $\eta\rightarrow 0$ \, .
      \end{itemize}

In order to obtain the light curve ${\cal I}(x)$ of the two--phase
flare model (again by integrating the emissivity distribution at each
$x$--coordinate in $y$--direction) the horn--shaped emissivity region
is divided into four distinct regions defined by (Fig.\,\ref{fig:6})
%
      \begin{eqnarray}
        x_1&=&\dsfrac{\sqrt{2}c\Ti}{2\gf^2}\dsfrac{1}{1+\sqrt{1+\kf
            \Ti/\Te}}\, ,\\
        x_2&=&\dsfrac{\sqrt{2}c\Ti}{2\gb^2}\dsfrac{1}{1+\sqrt{1+\kb
            \Ti/\Te}}\, ,\\
        x_3&=&\dsfrac{\sqrt{2}c\Te}{2\gf^2}\dsfrac{1}{1+\sqrt{1+\kf
        }}\, ,\\
        x_4&=&\dsfrac{\sqrt{2}c\Te}{2\gb^2}\dsfrac{1}{1+\sqrt{1+\kb
        }}\, .
        \end{eqnarray}
%
Then, using the auxiliary functions $H$, $g_{\rm b}$ and $g_{\rm f}$
defined in \S\,\ref{sec:horn_seg}, one can write the light curve of
the two--phase model in the form
%
        \begin{equation}\label{eq:twoplc}
          \begin{array}{l}
            {\cal I}(x)=j_0\sqrt{2}c\Te \times\\[2mm]
            \, \, \, \, \, \, \, \, \left\{
            \begin{array}{ll}
              H\left[g_{\rm b}(x),g_{\rm f}(x),
                0,\dsfrac{\Ti}{\Te},\theta_1,\eta_1\right] & 
              {\rm if}\,\, 0\leq x\leq x_1\\[10mm]
              \begin{array}{l}
                H\left[g_{\rm b}(x),\dsfrac{\Ti}{\Te},
                  0,\dsfrac{\Ti}{\Te},\theta_1,\eta_1\right]+\\[4mm]
                \eta_2 H\left[\dsfrac{\Ti}{\Te},g_{\rm f}(x),
                              \dsfrac{\Ti}{\Te},1,\theta_2,0\right]
              \end{array} & 
                          {\rm if}\,\, x_1< x\leq x_2\\[15mm]
                          \eta_2 H\left[g_{\rm b}(x),g_{\rm f}(x),
                            \dsfrac{\Ti}{\Te},1,\theta_2,0\right]
                          &  {\rm if}\,\, x_2< x\leq x_3\\[10mm]
                          \eta_2 H\left[g_{\rm b}(x),1,
                            \dsfrac{\Ti}{\Te},1,\theta_2,0\right]
                          &  {\rm if}\,\, x_3< x\leq x_4
            \end{array}
            \right.
          \end{array}
        \end{equation}
%
Figs.\,\ref{fig:6} and \ref{fig:7} show two examples of analytic
two--phase light curves, which only differ in the value of $\eta_2$
with $\eta_2=300$ in Fig.\,\ref{fig:6} and $\eta_2=101$ in
Fig.\,\ref{fig:7}, respectively.

\subsection{Normalized two-phase model}
\label{sec:norm}
      
As pointed out in \S\,\ref{sec:obs}, all observed time scales are
functions of $T/\Gamma^2$, where $T$ is a typical time scale in the
source frame, and $\Gamma$ is a typical Lorentz factor at which the
emitting region is moving. In order to make our model ``aware'' of
this, we introduce the normalized coordinate
%
      \begin{equation}\label{eq:ksi}
        \xi\equiv\dsfrac{x}{x_4}=\dsfrac{2\gb^2}{\sqrt{2}c\Te}
                                 (1+\sqrt{1+\kb})\, ,
      \end{equation}
%
and define
%
      \begin{equation}\label{eq:gamr}
        \gamma_{\rm b}\equiv\dsfrac{\gb}{\gf}
      \end{equation}
%
and
%
      \begin{equation}\label{eq:tb}
        t_{\rm i}\equiv\dsfrac{\Ti}{\Te}\, .
      \end{equation}
%
Thereby we can specify a normalized (both in intensity and in time)
flare model
%
      \begin{equation}\label{eq:normmodel}
        {\cal F}(\xi) = {\cal F}_0 \left\{ \begin{array}{lr}
            H[h_{\rm b}(\xi),h_{\rm f}(\xi),0,t_{\rm i},\theta_1,\eta_1] &
              {\rm if\,} 0\leq \xi\leq\xi_1\\[4mm]
            \begin{array}{l}
              H[h_{\rm b}(\xi),t_{\rm i},0,t_{\rm i},\theta_1,\eta_1]+\\[2mm]
              \eta_2 H[t_{\rm i},h_{\rm f}(\xi),t_{\rm i},1,\theta_2,0]
            \end{array} 
            & {\rm if\,} \xi_1< \xi\leq\xi_2\\[4mm]
            \eta_2 H[h_{\rm b}(\xi),h_{\rm f}(\xi),t_{\rm i},1,\theta_2,0] &
            {\rm if\,} \xi_2 < \xi\leq\xi_3\\[4mm]
            \eta_2 H[h_{\rm b}(\xi),1,t_{\rm i},1,\theta_2,0] &
                {\rm if\,} \xi_3 < \xi\leq 1
          \end{array}\right.
      \end{equation}
%
where the constant ${\cal F}_0$ is chosen such that the maximum value
of ${\cal F}(\xi)$ is $1$. The points $\xi_i$, $i=1, 2, 3$ are given
by
%
      \begin{equation}
        \xi_1\equiv \gamma_{\rm b}^2 t_{\rm i}
        \dsfrac{1+\sqrt{1+\kb}}{1+\sqrt{1+\kf t_{\rm i}}}\, ,
      \end{equation}
      \begin{equation}
        \xi_2\equiv t_{\rm i}
        \dsfrac{1+\sqrt{1+\kb}}{1+\sqrt{1+\kb t_{\rm i}}}\, ,
      \end{equation}
      \begin{equation}
        \xi_3\equiv \gamma_{\rm b}^2
        \dsfrac{1+\sqrt{1+\kb}}{1+\sqrt{1+\kf}}\, ,
      \end{equation}
%
and the auxiliary functions $h_{\rm b}(\xi)$ and $h_{\rm f}(\xi)$ are
obtained from $g_{\rm b}(x)$ and $g_{\rm f}(x)$, respectively:
%
      \begin{equation}
        h_{\rm b}(\xi) = \dsfrac{\kb\xi^2}{(1+\sqrt{1+\kb})^2}+
                         \dsfrac{2\xi}{1+\sqrt{1+\kb}}\, ,
      \end{equation}
      and
      \begin{equation}
        h_{\rm f}(\xi)=\dsfrac{\kf\xi^2}{\gamma_{\rm b}^4 
                       (1+\sqrt{1+\kb})^2}+
                       \dsfrac{2\xi}{\gamma_{\rm b}^2
                       (1+\sqrt{1+\kb})}\, .
      \end{equation}

The normalized two--phase model has the advantage that the absolute
intensity is no longer a parameter, and that it has embedded the
intrinsic degeneracies of $T/\Gamma^2$ which {\it must} be present in
the model due to the relativistic time contraction. Indeed, the
normalized two--phase model uses two parameter less in order to fit
the light curves. Therefore, we find that the fitting procedure is
always convergent.

\section{Fits of the analytic model to synthetic light curves}
\label{sec:fits}
      
The final goal of our analytic modeling of the light curve of a flare
is to extract unobservable physical parameters of the emitting region.
As a first step, we have validated our analytic model by comparing the
parameters extracted from fits\footnote{Due to the complex nature of
  the multidimensional parameter space of our analytic model
  (Eq.\,\eqref{eq:normmodel}), we have used genetic algorithms to fit
  it to the normalized light curves of our simulations.}
of synthetic light curves (computed from our hydrodynamic models) to
the parameters that we can directly obtain from the same hydrodynamic
models. The second step will be to apply our validated analytic model
to real observations, which is beyond the scope of this paper.
      
Table\,\ref{tab:3} shows the results of the fit of the analytic model
to the soft ($0.1$-$1$\,keV) light curves of all seven models. First,
we observe that the ratios $\gamma_{\rm b}^{\rm sim} \equiv \gb^{\rm
sim}/\gf^{\rm sim}$) and $\gamma_{\rm b} \equiv \gb/\gf$ are very
similar, but not exactly the same, i.e., there exists a correlation
between the values obtained from the simulations and the fit. Second,
the small values of the parameter $\kb$ indicate a lack of
acceleration of the reverse shock, i.e., the reverse shock moves at
almost constant velocity until it breaks out of the faster shell. This
is clearly seen in Fig.\,\ref{fig:9}, where the trajectory
(world--line) of the reverse shock (full thick line) is a straight
line between points $a$ and $d$, which correspond to the start of the
interaction and the time and place where the reverse shock breaks out
of the faster shell, respectively.  Third, the values of the parameter
$\kf$ imply that the forward shock accelerates, though its Lorentz
factor does not increase by more than about $30\%$ (model
\SF{10}{14}). From the simulations, we see that the forward shock
(dashed thick line in Fig.\,\ref{fig:9}) moves at approximately
constant velocity between points $a$ and $b$ the latter point
corresponding to the time and place where the forward shock catches up
with the reverse shock emerging from the leading edge of the slower
shell -- see Sec.\,\ref{sec:twop}). Afterwards, beyond point $b$, the
forward shock speeds up, because it breaks out of the slower shell.
Therefore, in the time interval between the formation of the forward
shock and the time when the emissivity associated to it ceases to
contribute to the light curve, there is a net acceleration of this
shock. This is also properly captured by our analytic model. Fourth,
on the one hand, considering the models having the same inertial mass
of the slower shell (rows 1--4 in Tab.\,\ref{tab:3}) the time $t_{\rm
b}$ necessary for the forward shock to break out of that shell,
relative to the total flare duration in the laboratory frame,
decreases with increasing inertial mass of the faster shell. On the
other hand, we find a similar correlation considering those models
where the inertial mass of the faster shell is fixed (rows 1, 5, 6
and 7), i.e., $t_{\rm b}$ is decreasing as the inertial mass of the
slower shell increases (see Sec.\,\ref{diss:hydro}).

      \begin{table*}
          \begin{center}{\large
            \begin{tabular}{|c||r|r|c|r|r|}
              \hline
              sim. & $\gamma_{\rm b}^{\rm sim}$ &
              $\gamma_{\rm b}$ &
              $\kb$ & $\kf$ & $t_{\rm b}$\\\hline\hline
              \SF{10}{10}  &
              $0.943$ & $0.832$ & $3.4\times 10^{-6}$ &
              $1.07$ & $0.429$\\\hline
              \SF{10}{07} &
              $0.940$ & $0.867$ & $1.2\times 10^{-4}$ &
              $1.68$ & $0.434$\\\hline
              \SF{10}{05} &
              $0.937$ & $0.861$ & $2.5\times 10^{-4}$ &
              $1.39$ & $0.445$\\\hline
              \SF{10}{14} &
              $0.944$ & $0.855$ & $6.4\times 10^{-4}$ &
              $2.35$ & $0.414$\\\hline
              \SF{07}{10} &
              $0.946$ & $0.913$ & $7.2\times 10^{-5}$ &
              $1.40$ & $0.567$\\\hline
              \SF{05}{10} &
              $0.948$ & $0.887$ & $1.1\times 10^{-4}$ &
              $1.95$ & $0.579$\\\hline
              \SF{14}{10} &
              $0.941$ &  $0.839$ & $1.09\times 10^{-4}$ &
              $1.34$ & $0.415$\\\hline
            \end{tabular}}
          \end{center}
          \caption{ Results of the fit of the normalized analytic
            model to the soft ($0.1$-$1$\,keV) light curves of all
            simulated models. $\gamma_{\rm b}^{\rm sim}$ is the ratio
            $\gb^{\rm sim}/\gf^{\rm sim}$ computed from the models
            (see Tab.\,\ref{tab:1}). The rest of the columns provide
            the other parameters obtained by fitting the normalized
            analytic model.}
          \label{tab:3}
      \end{table*}

      \begin{figure}
        \centering
        \includegraphics[scale=0.32]{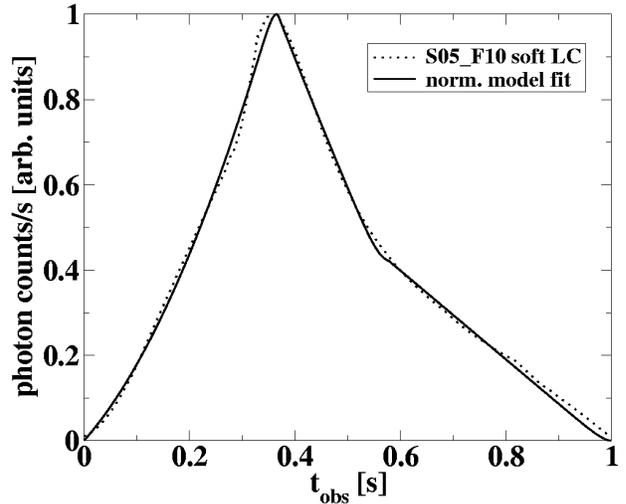}
        \caption{Soft light curve of model \SF{05}{10} (dotted line),
          and the best fit of the normalized analytic model (full
          line). }
        \label{fig:8}
      \end{figure}

\section{Space time emissivity distribution}
\label{sec:spacetime}

      \begin{figure*}
        \centering
        \includegraphics[scale=0.6]{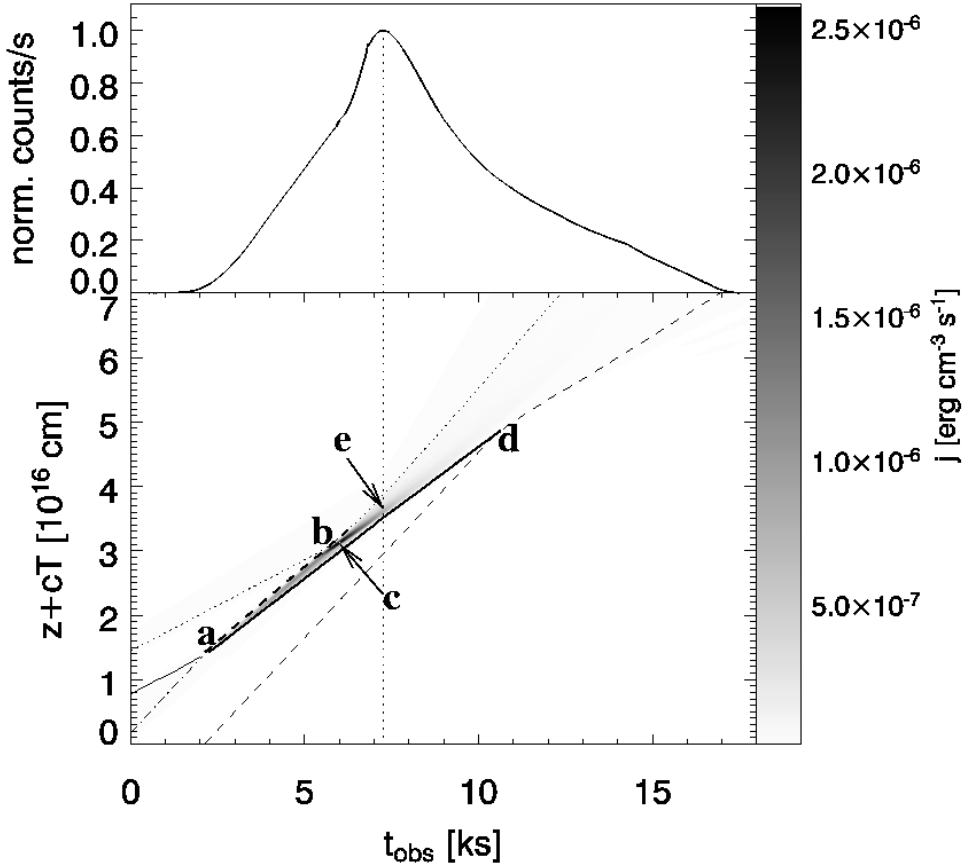}
    
        \caption{ The distribution of the emissivity in the observer
          frame in $xy$--coordinates in the soft photon band (lower
          panel), and the corresponding soft light curve (upper panel)
          for model \SF{05}{10}. The grey contours show the emissivity
          distribution on a linear scale. The full, dotted, dashed,
          and dot--dashed lines denote positions of the back edge of
          the slower shell, the front edge of the slower shell, the
          back edge of the faster shell, and the front edge of the
          faster shell, respectively. The full thick and dashed thick
          lines show the trajectories of the reverse and forward
          shock, respectively.  }
        \label{fig:9} 
      \end{figure*}
      
From Fig.\,\ref{fig:9}, which shows the soft light curve (upper panel)
and a grey contour plot of the emissivity distribution in the observer
frame using the $xy$--coordinates (lower panel) for model \SF{05}{10},
one sees that the emissivity distribution has the shape expected from
the analytic model, i.e., a horn--like shape starting at point $a$
where the internal shocks form. The region then widens as the shocks
propagate through the shells. The bulk of the radiation is produced
relatively early in the evolution (up to point $c$ in
Fig.\,\ref{fig:9}, which corresponds to the moment of maximum
emissivity), i.e., the subsequent formation of rarefaction waves does
not have any influence on the observed light curve (see, however,
\S\,\,\ref{diss:hydro} for a discussion of the possible influence on
later collisions). We point out that there might be a difference
between the time (in the observer frame) when the light curve has a
maximum (point $e$), and the point when the emissivity has a maximum
(point $c$). This difference arises because the total observed
intensity at a given time is the result of an integration of the
emissivity over the emitting region. Thus, the maximum observed
intensity can result from a region where the emissivity is not
maximal, but whose emitting volume is sufficiently large. In case of
model \SF{05}{10} (Fig.\,\ref{fig:9}) the maximal observed intensity
is delayed with respect to the maximum of the emissivity. This
influences the shape of the light curve, because prior to the maximum,
at $t_{\rm obs}\approx 6$\,ks, we see a \emph{kink} in the light curve
where its slope steepens.  When the emissivity maximum (in the
observer frame) happens close to, or even after the intensity maximum,
no such kink will appear (see Figs.\,\ref{fig:6} and \ref{fig:7},
which illustrate precisely these two cases).

\section{Discussion}
\label{sec:discussion}
    
We have performed a detailed study of two--shell collisions in the
framework of the internal shock model using relativistic hydrodynamic
simulations. The non--thermal radiation emitted by shocked plasma is
consistently coupled to the hydrodynamics (including the radiation
back--reaction on the dynamics) in our numerical scheme. Our findings
can be applied to extract physical parameters of the emitting regions
of blazar flares. However, the method of extracting physical
parameters out of synchrotron light curves might as well be applied to
other astrophysical scenarios where internal shocks occur (e.g.,
gamma--ray bursts).

\subsection{Hydrodynamic evolution of the shell interaction}
\label{diss:hydro}
      
The properties of the interaction common to all our models are the
existence of three main phases of evolution (Figs.\,\ref{fig:1} and
\ref{fig:2}): (i) the pre--collision phase where the front edges of
the shells pile up matter from the external medium and are heated due
to their interaction with the latter, (ii) the collision phase where
the internal shocks form and propagate through the shells, and (iii)
the post--collision or rarefaction phase where the reverse shock has
broken out of the faster shell creating rarefaction waves which expand
the merged shell and cause its structure to become multi--peaked.
Although this multi--peaked structure does not have any influence on
the observed flare, one has to bear in mind that any shell colliding
subsequently with the merged shell will encounter a non--uniform
structure, which will probably cause the forward shock, originating
after the new shell collision, to undergo a much more complicated set
of non--linear wave interactions than in the case of initially uniform
shells. This will yield a more complex light curve profile. We also
point out that the insensitivity of the light curve profiles to the
large rest--mass density and specific energy variations in the merged
shell, may probably prevent one from extracting from any simple {\em
  one-zone} model accurate information about these hydrodynamic
variables from fits of the observed flare light curves.  This lack of
sensitivity to the rest--mass density profile comes from the fact that
most of the steep density gradients in the merged shell are either
weak shocks or no shocks at all. Hence, no significant emission can be
originated in such steep density variations and the resulting light
curve remains uninfluenced.  Furthermore, when assuming an
approximately uniform proportionality between the rest--mass energy
and the magnetic field energy (i.e., the ratio $B^2/\rho$ being
approximately uniform) accurate values of the magnetic field strength
are hardly to be inferred from a flare light curve, too.
      
We have also computed the instantaneous efficiency of conversion of
kinetic energy into internal energy of the fluid. As can be seen from
Fig.\,\ref{fig:3}, the efficiency rises initially, then remains
constant for a period of time during which the front edge of the
faster shell propagates through the rarefaction caused by the back
edge of the slower shell, and then rises again when the internal
shocks form. The efficiency during the period when most of the
radiation is observed is larger than the analytically predicted
value. Afterwards, it decreases and reaches the analytically predicted
values.
      
For a fixed inertial mass of the slower shell, the Lorentz factor of
the reverse and forward shocks grows as the inertial mass of the
faster shell increases. Conversely, fixing the inertial mass of the
faster shell, the Lorentz factor of the shocks decreases with
increasing inertial mass of the slower shell. These correlations are
expected, because increasing the inertial mass of the slower shell any
incoming (faster) shell will supply less inertia (relative to that of
the slower shell), i.e., the two internal shocks will propagate at
smaller velocities. On the other hand, increasing the inertia of the
faster shell will result in larger shock speeds, because the target
shell does not decelerate the faster shell so efficiently.
      
The parameter $t_{\rm i}$, which is approximately the ratio of the
time it takes the forward and reverse shocks to cross the slower and
faster shell, respectively, depends on the inertial masses of the
shells, too. We find that $t_{\rm i}$ decreases with increasing
relative inertia of the faster shell (with respect to the slower one),
because with increased inertia it is much easier for the forward shock
to propagate through the slower shell and exiting from its front edge
earlier than the reverse shock, which propagates into the heavier,
faster shell.

\subsection{Properties of the analytic model}
\label{diss:analytic}
      
As described in \S\,\ref{sec:analytic}, we have introduced the
observer--xy coordinate system in order to simplify the relationship
between the observed radiation and the space time evolution of the
emissivity. By making assumptions about the properties of the
emissivity distribution, we are able to construct an analytic model
which depends on parameters of the internal shocks, namely the shock
Lorentz factors and the shock propagation time scales. Our model can
be applied to non--relativistic as well as ultra-relativistic shocks,
and even allows for a (constant) acceleration of the shocks. For the
moment, it does not take into account any spectral information, but we
intend to improve our model in order to be able to interpret
multi--frequency observations which may provide further restrictions
on the physical parameters of the emitting regions.  We have validated
the analytic model by comparing its fitted parameters (using synthetic
light curves) with their values obtained directly from the simulations
which provide the synthetic light curves.

\subsection{Synthetic flares and the analytic model fitting}
\label{diss:flares}
      
We find that all flares, once normalized, look qualitatively the same.
This suggests that the exact shape of the flare depends mostly on the
shock Lorentz factors and shock propagation time scales while it is
rather insensitive to the distribution of the rest--mass density or
specific energy within the emitting region (see \S\,\ref{diss:hydro}).
We have fitted the analytic model (using a simple genetic algorithm)
to synthetic light curves and have achieved satisfactory results. The
deviations in the parameters between the simulation and the fit
(Tab.\,\ref{tab:3}) are due to the simplifications of the analytic
model, e.g., the assumption that the emissivity is constant between
internal shocks (this is not exactly the case, as can be seen in
Fig.\,\ref{fig:9}). However, the general shape of the emissivity
distribution (Fig.\,\ref{fig:9}) does match our assumptions. This
implies that one might fit the model parameters using the observed
X-ray light curves of blazar flares (e.g., Mrk\,421) and recover shock
parameters, like e.g., the ratio of Lorentz factors between the
reverse and forward shock, the propagation time scales through the
shells, and the kinematic evolution (acceleration) of the shock
fronts. However, here the quality of the observations might play a
crucial role, since the analytic model is very sensitive to changes in
the flare shape, as can be seen from Figs.\,~\ref{fig:6} and
\ref{fig:7}.

We have produced synthetic observations in soft and hard energy bands
(Fig.\,\ref{fig:4}). A correlation between the peak photon counts and
the initial rest mass density of the shells has been found.
Currently, we are performing a more detailed parameter study aiming to
confirm this suspected correlation for cases of different shell
Lorentz factors and densities. The peak in the hard band is observed
several hundreds of seconds before the one in the soft band. However,
the exact value of the lag between the hard and soft peaks will be
different in the case of varying shock strengths and velocities (which
strongly depend on the shell properties). Additionally, the possible
acceleration of particles inside the emitting region is not included
in our model. Instead, particles are injected into the plasma with a
given energy distribution, as acceleration timescales are expected to
be shorter than the dynamical timescale. Note, however, that
acceleration may be the origin of some of the features (sign of the
delay) observed in temporal evolution of these sources (e.g.,
\cite{KRM98}~1998).

Observationally, the lags between the peaks in the soft and hard bands
are still a matter of controversial debate. The emission of the soft
X-rays can be well correlated with that of the hard X-rays lagging it
by 3$-$4\,ks (\cite{TA96} 1996, 2000, \cite{ZH99} 1999, \cite{MA00}
2000, \cite{KA00} 2000, \cite{FO00} 2000).  However, significant lags
of both signs were detected from several flares (\cite{TA01} 2001).
The lags between the soft and hard peaks may depend on the assumed
value of the parameter $\alpha_{B} = (B^2/8\pi)/(\rho \varepsilon)$
which sets the strength of the magnetic field in the emitting region.
As in \MA04 this parameter has been chosen such that the field
strength resulting from the hydrodynamic evolution yields values of
$\sim 0.1$\,G (the value of $\alpha_{B}$ is the same for all our
models). This field strength lies in the ballpark of the values
inferred from detailed fits of SED curves of several sources (see,
e.g., \cite{TA01}\,2001). We point out that even for the same source
there is no consensus on the exact value of the magnetic field
strength. For example, \cite{TA01}\,(2001) report $B \sim 0.14\,$G for
PKS\,2155-304, while is inferred $B \sim 1.5\,$G from
\cite{Ch99}\,(1999) for the same source (taking the Doppler factor to
be $\delta = 28$; \cite{TA01} 2001).
      
Observations of PKS\,2155-304 \cite{ED01} (2001) using \xmm suggest
that previous claims of soft lags with time scales of $\sim$\,hours
might be an artifact of the periodic interruptions of the low--Earth
orbits of the satellites every $\sim 1.6$\,hours.  This claim was
questioned by \cite{ZC04} (2004) who show that, although periodic gaps
introduce larger uncertainties than present in evenly sampled data,
lags on time scales of hours cannot be the result of periodic gaps.
Large flares with time scales of $\sim 1$\,day were detected with
temporal lags of less than 1.5\,hours between X-ray and TeV energies
(for Mrk\,421 see \cite{TA00} 2000).
      
Recently, very detailed studies of Mrk\,421 with \xmm (\cite{BR03}
2003, 2005; \cite{Retal04}\,2004) clearly show that the source
exhibits stronger variability when it is brighter, that the cross
correlation function appears to change ``continuously'', and depending
on the length of the observing window and the actual activity state of
the source they find periods with positive, negative or no lags, but
also periods of weak correlations between the soft and hard energy
bands.  Here we propose that the apparent sign of the lags is the
result of the particular arrangement of a number of emitting regions,
i.e., the resulting spectral properties will depend strongly on the
time scales of their emission, as well as their distances and relative
velocities.

\begin{acknowledgements}
  All computations were performed on the IBM-Regatta system of the
  Rechenzentrum Garching of the Max-Planck-Society. PM acknowledges
  support from the Special research Area SFB~375-95 on Astro-Particle
  Physics of the German Science Foundation. MAA is a Ram\'on y Cajal
  Fellow of the Spanish Ministry of Education and Science.  MAA
  acknowledges the partial support of the Spanish Ministerio de
  Ciencia y Tecnolog\'{\i}a (AYA2001-3490-C02-C01).
\end{acknowledgements}


\end{document}